%% file: main.tex
\documentclass[sigconf]{acmart}

\usepackage{soul}
\usepackage{algorithm2e}
\usepackage{hyperref}
\usepackage{balance}
\usepackage{todonotes}
\usepackage{nameref}
\usepackage{booktabs}
\usepackage{tabularx}
\usepackage{multirow}
\usepackage{stfloats}

\newcolumntype{L}[1]{>{\raggedright\arraybackslash}m{#1}}
\newcolumntype{Y}{>{\raggedright\arraybackslash}X}

\definecolor{oxfordblue}{rgb}{0.0, 0.13, 0.28}
\definecolor{harvardcrimson}{rgb}{0.79, 0.0, 0.09}
\definecolor{dartmouthgreen}{rgb}{0.05, 0.5, 0.06}
\definecolor{princetonorange}{rgb}{1.0, 0.56, 0.0}
\definecolor{yaleblue}{rgb}{0.06, 0.3, 0.57}
\definecolor{usccardinal}{rgb}{0.6, 0.0, 0.0}
\definecolor{uclablue}{rgb}{0.33, 0.41, 0.58}
\definecolor{msugreen}{rgb}{0.09, 0.27, 0.23}
\definecolor{cornellred}{rgb}{0.7, 0.11, 0.11}
\definecolor{pomegranate}{RGB}{192, 57, 43}
\definecolor{anti-pomegranate}{RGB}{43,178,192}
\definecolor{alizarin}{RGB}{231, 76, 60}
\definecolor{anti-belize}{RGB}{185, 41, 56}
\definecolor{belize}{RGB}{41, 128, 185}
\definecolor{peter}{RGB}{52, 152, 219}
\definecolor{green}{RGB}{22, 160, 133}
\definecolor{anti-green}{RGB}{160,22,118}
\definecolor{turquoise}{RGB}{26, 188, 156}
\definecolor{pumpkin}{RGB}{211, 84, 0}
\definecolor{anti-pumpkin}{RGB}{0,22,211}
\definecolor{carrot}{RGB}{230, 126, 34}
\definecolor{wisteria}{RGB}{142, 68, 173}
\definecolor{anti-wisteria}{RGB}{99,173,68}
\definecolor{amethyst}{RGB}{155, 89, 182}
\definecolor{nephritis}{RGB}{39, 174, 96}
\definecolor{anti-nephritis}{RGB}{174,39,117}

\newcommand{\cmt}[1]{\ignorespaces}

\SetKwInput{KwInput}{Input}
\SetKwInput{KwOutput}{Output}
\SetKwInput{KwName}{Name}

\usepackage{mathtools}

\AtBeginDocument{%
  \providecommand\BibTeX{{%
    \normalfont B\kern-0.5em{\scshape i\kern-0.25em b}\kern-0.8em\TeX}}}

\setcopyright{acmcopyright}
\copyrightyear{2024}
\acmYear{2024}
\acmDOI{XXXXXXX.XXXXXXX}


\begin{document}


\title[AI Assistance for UX: A Literature Review]{AI Assistance for UX: A Literature Review Through Human-Centered AI}


\author{Yuwen Lu}
\affiliation{%
  \institution{University of Notre Dame}
  \city{Notre Dame}
  \state{IN}
  \country{USA}}
\email{ylu23@nd.edu}

\author{Yuewen Yang}
\affiliation{%
  \institution{Cornell University}
  \city{New York}
  \state{NY}
  \country{USA}}
\email{yy2228@cornell.edu}

\author{Qinyi Zhao}
\affiliation{%
  \institution{University of Washington}
  \city{Seattle}
  \state{WA}
  \country{USA}}
\email{qyzhao@uw.edu}

\author{Chengzhi Zhang}
\affiliation{%
  \institution{Georgia Institute of Technology}
  \city{Atlanta}
  \state{GA}
  \country{USA}}
\email{czhang694@gatech.edu}

\author{Toby Jia-Jun Li}
\affiliation{%
  \institution{University of Notre Dame}
  \city{Notre Dame}
  \state{IN}
  \country{USA}}
\email{toby.j.li@nd.edu}
\renewcommand{\shortauthors}{Anonymous Author et al.}

\begin{abstract}

Recent advancements in HCI and AI research attempt to support user experience (UX) practitioners with AI-enabled tools. Despite the potential of emerging models and new interaction mechanisms, mainstream adoption of such tools remains limited. We took the lens of Human-Centered AI and presented a systematic literature review of 359 papers, aiming to synthesize the current landscape, identify trends, and uncover UX practitioners' unmet needs in AI support. Guided by the Double Diamond design framework, our analysis uncovered that UX practitioners' unique focuses on empathy building and experiences across UI screens are often overlooked. Simplistic AI automation can obstruct the valuable empathy-building process. Furthermore, focusing solely on individual UI screens without considering interactions and user flows reduces the system's practical value for UX designers. Based on these findings, we call for a deeper understanding of UX mindsets and more designer-centric datasets and evaluation metrics, for HCI and AI communities to collaboratively work toward effective AI support for UX.

\end{abstract}





\settopmatter{printfolios=true}
\maketitle

\input{sections/1-introduction}
\input{sections/2-related-work}
\input{sections/3-literature-selection}
\input{sections/4-0-analysis}
\input{sections/4-1-discover}
\input{sections/4-2-define}
\input{sections/4-3-develop}

\input{sections/4-4-deliver}
\input{sections/4-5-datasets}

\input{sections/4-6-general-ai-models}
\input{sections/5-discussion}

\input{sections/6-conclusion}
\input{sections/7-limitations}

\balance
\bibliographystyle{ACM-Reference-Format}
\bibliography{references}

\clearpage
\appendix
\input{sections/z-appendix}

\end{document}

%% file: sections/1-introduction.tex
\section{Introduction}
Advancements in Artificial Intelligence (AI) enabled applications in numerous sectors, with the user experience (UX) industry being a notable potential beneficiary. AI models can facilitate processes that involve various data modalities, ranging from text-based affinity diagrams~\cite{goldman_quad_2022, borlinghaus_comparing_2021} and user interface (UI) development codes~\cite{beltramelli_pix2code_2017, feng_guis2code_2021} to image-based UI screenshots~\cite{leiva_describing_2022, wang_screen2words_2021, zhao_guigan_2021}. The enhancements of language-based and multi-modal AI models have expanded the possibilities of applications in UX design and research~\cite{dhinakaran_survey_nodate, di_fede_idea_2022, kim_cells_2023}. Notably, the impressive capabilities of large-language models (LLMs) further promoted AI adoption in real applications~\cite{dhinakaran_survey_nodate}. Diffusion-based, text-to-image generative AI such as Stable Diffusion~\cite{rombach_high-resolution_2022} and Midjourney\footnote{https://www.midjourney.com/} also opens up new avenues for creative professionals to utilize AI in their work~\cite{verheijden_collaborative_2023, wei_boosting_2023}.

However, creating usable, effective, and enjoyable AI-enabled experiences for UX practitioners remains challenging~\cite{yang_re-examining_2020}. A technology-driven mindset, prevalent in AI communities, can lead to applications that are driven by the latest technology, but do not necessarily address UX practitioners' unique goals such as empathy-building. Furthermore, the fluid, nonlinear UX methodologies~\cite{gray_its_2016} are not the same as logical, computational thinking and can be hard to grasp for AI researchers. The lack of insight into designer workflow and practices can create challenges for AI research to create effective and seamless support for UX professionals.

Not all UX processes are desired to be delegated to AI~\cite{marathe_semi-automated_2018, lubars_ask_2019}, leading to concerns about the diminished empathy of the designer when valuable research processes become automated. Such concerns question the real-world efficacy of these AI models in providing meaningful UX support. Early research prototypes on AI-enabled design support systems have received positive feedback in user studies~\cite{cheng_play_2023, hegemann_cocolor_2023, rietz_cody_2021, gebreegziabher_patat_2023}. At the same time, unique data modalities, user needs, and workflows in UX also created new practical challenges for AI researchers to tackle~\cite{li_screen2vec_2021, rietz_cody_2021, gebreegziabher_patat_2023, wang_screen2words_2021}. 

The field of human-centered AI (HCAI) provides valuable perspectives for investigating the current gap and future risks in AI for UX support. HCAI sits at the intersection of AI and Human-Computer Interaction (HCI) and embraces the human-centered philosophy. It aims to ensure that AI systems align with human values and mitigate potential harms to individuals, communities, and societies~\cite{shneiderman_human-centered_2022}. As AI models integrate into more real-world applications, it becomes imperative to prioritize human-centered design and research principles in AI adoption. Researchers in HCAI have investigated useful design metaphors and paradigms for AI systems~\cite{yang_unremarkable_2019, shneiderman_human-centered_2022}.

In this work, we conducted a systematic literature review (SLR) through the lens of HCAI and analyzed the state of technical and system research in AI assistance for UX practitioners. We outline the role of AI in different phases of UX practices using the classic Double Diamond design framework~\cite{british_design_council_double_nodate}. Our SLR sought to understand AI's current technical capabilities with UX-related tasks and map out the rapidly expanding design space of AI for UX support. Our general goal is to pinpoint opportunities for both HCI and AI communities, to identify the critical needs of UX professionals, and to find common ground between UX practices and frontier academic AI research. Thus, we define our research questions as follows.

\label{intro:research-questions}

\begin{enumerate}
    \item What capabilities do the latest AI models possess for different UX-related tasks?
    \item Regarding UX practitioners' needs and preferences for AI assistance, what insights have been revealed from past research?
    \item What are the gaps between existing empirical studies and opportunities for future AI research and interactive system development?
\end{enumerate}

Through our SLR with 359 papers, we found that past work has a higher focus on technology-driven approaches than human-centered investigations. Our analysis underscored the contrast between AI's data-driven nature and the human-centric philosophy of UX. Building on this, our study maps existing research onto the Double Diamond framework~\cite{british_design_council_double_nodate}, identifying key technical capabilities of AI in UX (Section~\ref{section:analysis-general}) and underscoring overlooked areas such as empathy-building and enhancing user experiences across multiple UI screens (Section~\ref{section:general-ai-models}). The UX industry can also benefit from embracing data-driven strategies to capture feedback from ever-expanding user bases. We emphasize the need for a deeper understanding of UX methodologies and goals, the expansion of quantitative UX metrics, and careful consideration of AI delegability based on existing Human-Centered AI frameworks~\cite{lubars_ask_2019}. This work aims to offer valuable insights and direction for future research to the HCI, UX, and AI communities, highlighting the potential of this promising interdisciplinary, translational research domain.

%% file: sections/2-related-work.tex
\section{Background and Related Work}
\subsection{UI/UX Design and Support Tools}
\label{section:ui-ux-design-and-support-tools}
UI/UX as a profession has established its status in both the tech industry and academia over the past decades. Nielson estimated that the population of UX professionals worldwide grew from about 1,000 to 1 million between 1983 and 2017. It is also estimated that in 2050, the number will increase by another 100-fold to 100 million~\cite{nielsen_100-year_2017}. UX practitioners aim to create products and experiences that are user-friendly, enjoyable, and effective. They often try to understand target users' needs through human-centered methodologies, e.g. contextual interviews, and iteratively prototype their design solutions and elicit feedback from users. Such a process is well captured in British Design Council's Double Diamond framwork~\cite{british_design_council_double_nodate}. Through two divergent-convergent processes, UX practitioners brainstorm and select particular aspects of an issue to tackle, then iteratively prototype a few potential solutions and finalize on one through user feedback.

Numerous support tools have been developed for UX design. From early HCI research, the SILK system was one of the first no-code designer-support UI prototyping tools~\cite{landay_silk_1996}. Later, Sketch and Figma are among the most popular tools for UX prototyping. More related to the early exploratory phases, platforms such as Miro, Mural, and FigJam are created for UX professionals to organize ideas, conduct brainstorming, or qualitatively analyze user data. Evaluation platforms such as UserTesting\footnote{https://www.usertesting.com/} and Maze\footnote{https://maze.co} provide support for conducting user evaluations, while researchers also investigated automated design testing~\cite{deka_zipt_2017} and remote user testing~\cite{martelaro_needfinding_2017}. Notably, design systems such as Google Material Design and Apple Human Interface Guidelines also provided tools to help designers create user-friendly, consistent, and accessible UIs.

Recently, we have witnessed an increase in AI integration into design support tools in both academia and industry. In academia, many researchers have been exploring AI-enabled support tools for UX practitioners~\cite{li_screen2vec_2021,  sermuga_pandian_uisketch_2021, lu_bridging_2022, knearem_exploring_2023}. In the industry, design tools like Uizard\footnote{https://uizard.io/} and Framer\footnote{https://www.framer.com/ai} have rolled out AI features to generate UI screens from natural language descriptions. Figma also recently acquired Diagram\footnote{https://diagram.com/}, a startup that previously focused on AI-enabled Figma plugins, and started to roll out AI features in their tool. However, the UX industry embodies a human-centered principle, which is inherently different from the technology-first mindset prevalent in AI communities. This has created friction in designing better AI experiences~\cite{yang_re-examining_2020} as well as creating effective AI support for UX practitioners~\cite{lu_bridging_2022}. We have yet to observe any of these AI-enabled tools become mainstream and adopted by a significant portion of the UX industry. This might reflect a ``research-practice gap'' that is common across HCI research~\cite{norman2010research}, solving which requires more translational research and resources to fulfill the needs of practitioners~\cite{colusso2017translational}.

\subsection{Human-Centered AI}

Human-Centered AI (HCAI) is an emergent interdisciplinary research field that bridges AI and HCI. HCAI embraces the human-centered philosophy and takes a humanistic and ethical view towards the latest AI technology: how to enhance humans rather than replace them~\cite{xu_toward_2019}. Researchers in HCAI have predicted that by embracing a human-centered future, the AI community's impact will likely grow even greater~\cite{shneiderman_human-centered_2022}. 

The primary research focuses of HCAI include: (1) improving AI-driven technology to better augment human needs, (2) identifying design methodologies for safe and trustworthy AI systems, and (3) understanding and safeguarding the impact of AI on individuals, communities, and societies~\cite{xu_toward_2019, shneiderman_human-centered_2022}. In this work, we investigate AI support for UX practitioners through the lens of HCAI, proposing our research questions (see the \nameref{intro:research-questions} section) based on the research focuses above. We refer to past research in HCAI, including Principles of Mixed-Initiative Interfaces~\cite{horvitz_principles_1999}, Guidelines for Human-AI Interaction~\cite{amershi_guidelines_2019}, and books on Human-Centered AI~\cite{shneiderman_human-centered_2022}. Particularly, we balance our analysis on both the \textit{technical} and \textit{design} aspects, seeking to understand existing AI models' capabilities in UX tasks, as well as practitioners' needs for automation in current methodologies and practices.

\subsection{Literature Review in AI Support for UI/UX design}

Past literature review studies in computing and HCI have successfully identified trends and gaps and proposed new research directions in different specific domains~\cite{dell_ins_2016, dillahunt_sharing_2017, lopez_awareness_2017, pater_standardizing_2021, stefanidi_literature_2023}. We consider the call for more literature review studies in HCI, CSCW, and Ubicomp~\cite{lopez_awareness_2017} and specifically look at the emerging field of AI for UI/UX design support.

While many researchers conducted general investigations on this topic~\cite{lu_bridging_2022, knearem_exploring_2023, isgro_ai-enabled_2022, liao_framework_2020, grigera_ai_2023}, only 3 papers used systematic literature review by the time we conducted this study. Malik et al. reviewed 100 papers and analyzed the deep learning approaches that have been utilized to support UI/UX design work~\cite{malik_reimagining_2023}. Their analysis results revealed potential for cross-platform datasets, more advanced UI generation models, and a centralized deep-learning-based design automation system. 

In addition, Abbas et al.~\cite{abbas_user_2022} analyzed 18 papers in this field and analyzed UX designers' current challenges in incorporating ML in their design process. Their results showed that most ML-enabled UX design tools fail to be integrated in practical settings. They argued the need to build support tools by considering existing design practices, rather than simply based on existing ML models' capabilities. Interestingly, the paper did not distinguish designing with ML support (the focus of our paper) from designing ML-involved systems and experience (i.e., AI as a design material, outside of our scope). Many of their summaries and discussions were centered around the need for designers' understanding of ML, which is beyond the scope of our analysis.

In 2022, Stige et al.~\cite{stige2023artificial} conducted a literature review on 46 articles in this field to analyze how AI is currently used in UX design (namely, user requirement specification, solution design, and design evaluation) and potential future research themes. Compared to their analysis sample (N=46), our sample was more comprehensive (N=359) and up-to-date (conducted in 2023), resulting in a more complete analysis of the recent empirical and technical research landscape (Section~\ref{section:analysis-general}). In addition, by mapping previous research into the four phases of the Double Diamond framework, we revealed more details regarding AI's involvement in UX research and design activities. Our analysis also uncovered more in-depth differences between AI and UX communities' mindsets and pointed out meaningful gaps to bridge for future research (Section~\ref{section:discussion}).

%% file: sections/3-literature-selection.tex
\section{Literature Review Method}
\label{sec:method}
To address our research questions (see \nameref{intro:research-questions}), we conducted a \textit{systematic literature review (SLR)} of papers in relevant research fields. SLRs are designed to help understand and interpret a large volume of information, to explain ``what works'' (i.e., current landscape) and ``what should work'' (i.e., potential gaps and future directions) in a given field. The ``systematic'' aspect of SLR focuses on identifying all research that addresses a specific question to conduct a balanced and unbiased summary~\cite{nightingale_guide_2009}. We followed previous guidelines on conducting SLRs~\cite{xiao_guidance_2019, nightingale_guide_2009} and referred to previous SLR studies in adjacent fields to form our methods~\cite{kaluarachchi_systematic_2023, dillahunt_sharing_2017, pater_standardizing_2021, wohlin_guidelines_2014}.

\begin{figure*}[htbp]
    \centering
    \includegraphics[width=\textwidth]{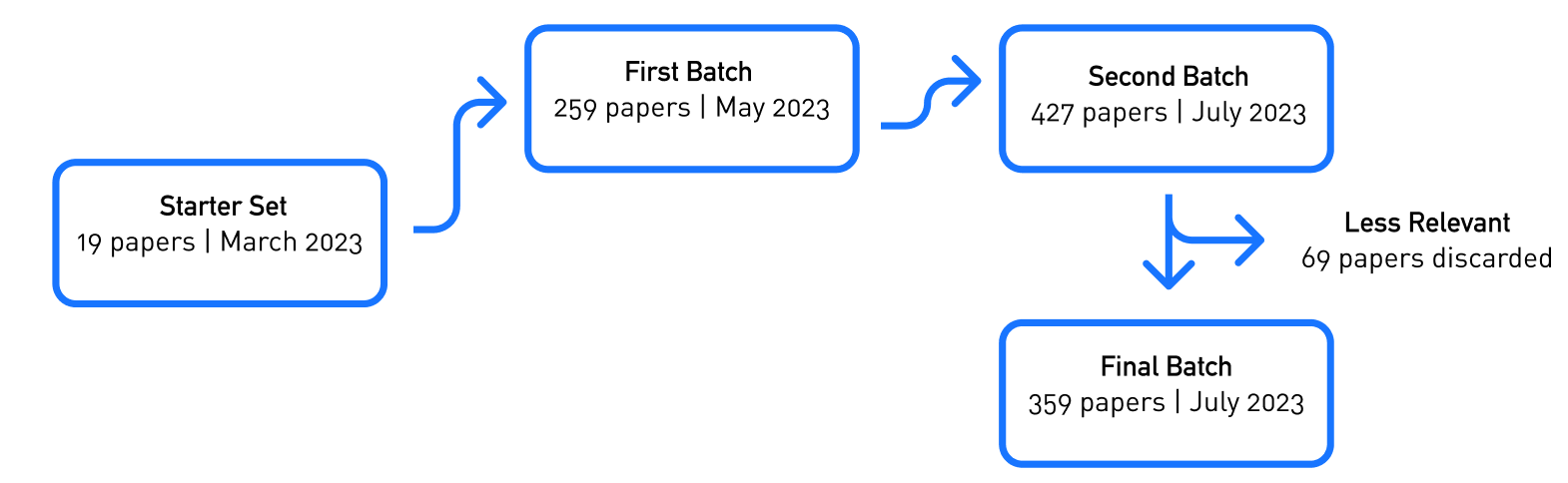}
    \caption{Adapted PRISMA statement of our snowball sampling process.}
    \label{fig:snowball_process}
\end{figure*}

We used \textit{snowball sampling}, a widely adopted literature search strategy, to select our literature sample\footnote{More explanation of our rationale to use snowball sampling can be found in Appendix~\ref{sec:appendix}.}. It begins with a starter set of a few relevant papers, then iteratively includes related papers that were cited by, or cited, papers in the starter set (i.e., the backward and forward snowballing processes)~\cite{wohlin_guidelines_2014}. Google Scholar was used as our primary search engine, as it is one of the largest online academic search engines, and is commonly used in literature review studies~\cite{wohlin_guidelines_2014, xiao_guidance_2019, siddaway_how_2019, cheng_sharing_2016}. We did not restrict the publication venues to reduce bias and get a diverse sample across disciplines~\cite{nightingale_guide_2009}. We depict our process in Fig. ~\ref{fig:snowball_process} by following an adapted version of the PRISMA statement~\cite{moher_preferred_2009}. Below, we detail our literature selection process, including our inclusion/exclusion criteria, the selection of a starter set, and iterative backward and forward samplings.

\subsection{Inclusion/Exclusion Criteria}

In line with our research scope, we included papers that satisfy both of the following criteria:
\label{section:selection-criteria}
\begin{itemize}
    \item providing support for methodologies or artifacts in UI/UX design and research,
    \item incorporating the use of artificial intelligence for such support.
\end{itemize}

We referred to articles regarding UX design and research practices~\cite{rosala_discovery_2020, farrell_ux_2017, pernice_affinity_2019, rosala_how_2022} to inform our selection process against the first criteria. Specifically, we used the Double Diamond design framework~\cite{british_design_council_double_nodate} to map out opportunities for AI support in the UX workflow, similar to past studies in this domain~\cite{yang_re-examining_2020}. We excluded papers that focused only on UI development without relevance to UI/UX design or research. 

As discussed in previous work, coming up with a precise, comprehensive definition of AI is hard, even within AI research communities~\cite{stone2022artificial}. It is even harder when considered in the HCI and UX contexts~\cite{yang_re-examining_2020} and is beyond the scope of our paper. We use the term as a reference to a suite of computational techniques generally considered in the domain of AI, from neuron-network-based deep learning models to statistical, machine learning approaches~\cite{russell2010artificial}. We excluded papers investigating the design of AI systems, often referred to as \textit{``AI as a design material''}~\cite{yang_re-examining_2020, yildirim_how_2022}. These papers often work on the designerly understanding of AI~\cite{liao_designerly_2023} and design processes that account for AI safety and accountability~\cite{moore_failurenotes_2023}. They focus more on \textit{\textbf{the design of AI}} instead of \textit{\textbf{supporting design with the helpf of AI}} (our focus). 

It is also noteworthy that our focus is specifically on AI adoption in UX support. While relevant, we do not aim to conduct a comprehensive literature review on creativity support tools, human-AI co-creation, or human-centered AI, given these are much broader research topics independent of our scope. However, we did draw inspiration from papers from these domains that do not fit our scope exactly and include them in our \nameref{section:discussion} section for better generalizability of our findings.

\subsection{Starter Set}

In the beginning, four researchers collaboratively searched for and filtered relevant papers using research search engines including Google Scholar and ACM Digital Library, based on our inclusion criteria defined in Section~\ref{section:selection-criteria}. When selecting our starter set, we followed previous work~\cite{nightingale_guide_2009} and aimed at the diversity of topics to minimize bias. Specifically, we also ensured to include a balanced set of papers addressing every phase of the Double Diamond process~\cite{british_design_council_double_nodate}. The four researchers frequently communicated and discussed in-depth during the selection process to ensure the representativeness and quality of our starter set. In the end, we included 17 papers related to the four Double Diamond phases (four, three, five, and five papers from discover, define, develop, and deliver, respectively). We also included two papers that investigate the same problem domain but do not specifically fit into any phase above to ensure representativeness and comprehensiveness. In all, our starter set consisted of 19 representative papers.

\subsection{Backward and Forward Sampling}

After selecting the starter set, we conducted two rounds of iterative sampling. In each iteration, both the papers that our sample cited (backward sampling of past papers) and the papers that cited our sample (forward sampling of later papers) were examined by four researchers. Researchers examined the full text of identified papers to determine their relevance, eligibility, and quality. A minimum of two researchers independently evaluated each paper and settled disagreements through discussions. Details of the iterations were depicted in Fig. \ref{fig:snowball_process}, following an adapted version of the PRISMA statement~\cite{moher_preferred_2009}. 

We stopped after the second snowballing iteration because we had already obtained a large sample (N=359) that is representative of the existing work in our domain. Also, in the second iteration, we observed that papers from the first iteration repeatedly appeared in papers of interest. In the analysis process, upon detailed examination, 68 papers were excluded due to their relative lack of relevance to our research questions. Our final sample contained a total of 359 papers, sourced from March to July 2023 (Fig~\ref{fig:snowball_process}). To the best of our knowledge, it is to date \textbf{the largest repository} of existing literature on the topic of AI for UX support compared to past literature reviews in this field~\cite{abbas_user_2022, malik_reimagining_2023, stige2023artificial}. 

%% file: sections/4-0-analysis.tex
\section{Analysis}
\label{section:analysis-general}
After all papers were selected and screened, the research team mapped their main topic into one of the four phases in the Double Diamond design framework~\cite{british_design_council_double_nodate}, a classic framework that comprehensively covers various activities in a design process. It has guided many previous academic research on UX design~\cite{gustafsson_analysing_2019, yang_re-examining_2020, ammarullah_design_2021}. It encapsulates the two divergent--convergent processes in design, where designers explore potential problems to address in the domain, then converge to main target issues; prototype a few potential solutions, and decide on the most effective one through testings and evaluations~\cite{british_design_council_double_nodate}. It should be noted that modern design processes are mostly iterative, so designers can go back and forth between different phases.

\begin{figure*}[htbp]
    \centering
    \includegraphics[width=0.7\textwidth]{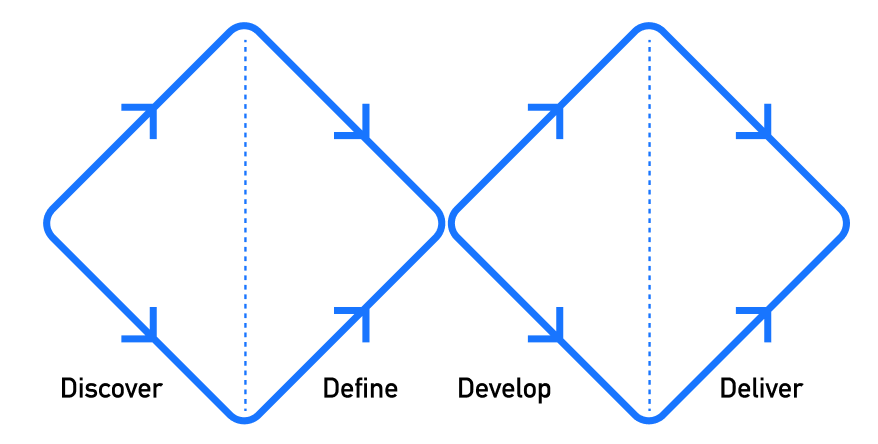}
    \caption{The Double Diamond design framework, proposed by the Design Council~\cite{british_design_council_double_nodate}.}
    \label{fig:double_diamond}
\end{figure*}

Given that we also focus on the technical feasibility of AI models in UX, two additional categories were also included: ``Datasets'' on UX-related datasets, and ``General AI Models'', about AI models that work with UX-related data and can be applied to more than one phase in the Double Diamond framework. When a paper fits more than one phase, we include it in the primary phase it belongs to.\looseness=-1

The papers in each phase were analyzed and discussed by at least two researchers. For each paper, based on our research questions and our human-centered AI perspective, we define the following seven aspects to focus on: 

\begin{enumerate}
    \item Research contribution type (according to~\cite{wobbrock_research_2016})
    \item Target problem/task
    \item Study/discussion of user needs
    \item Supporting empirical evidence from previous work (if any)
    \item AI model architecture and data modality
    \item Other important model aspects (e.g. user control, explainability)
    \item UX artifacts involved
\end{enumerate}

Researchers also took notes on meaningful information outside of these aspects. In a shared spreadsheet, researchers filled in information about the paper for the above aspects and discussed them for our analysis.

\begin{figure*}[htbp]
    \centering
    \includegraphics[width=\textwidth]{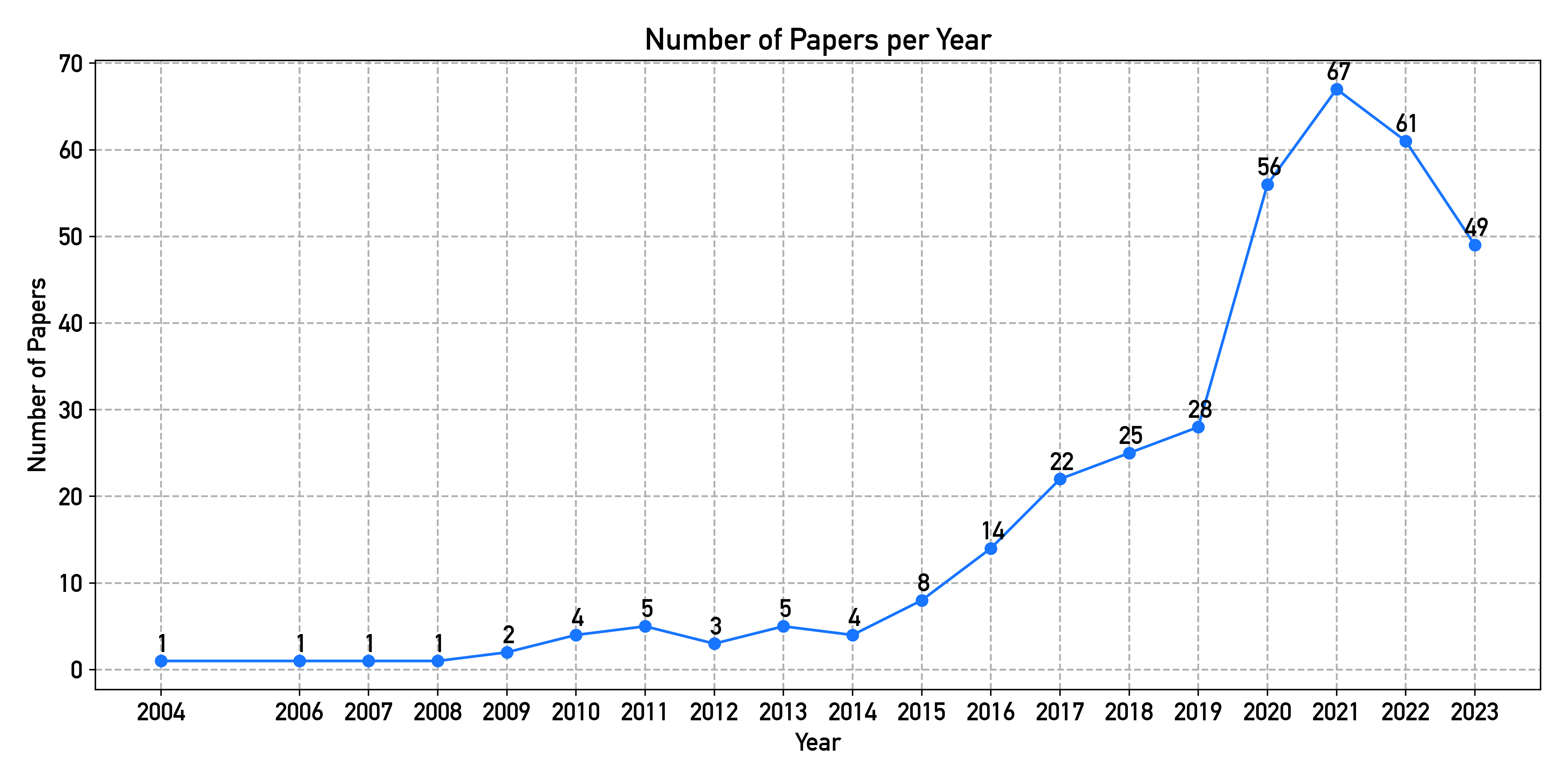}
    \caption{Trend of paper counts each year in our sample (the cut-off date of our search is July 2023).}
    \label{fig:paper_year}
\end{figure*}

\begin{figure*}[htbp]
    \centering
    \includegraphics[width=\textwidth]{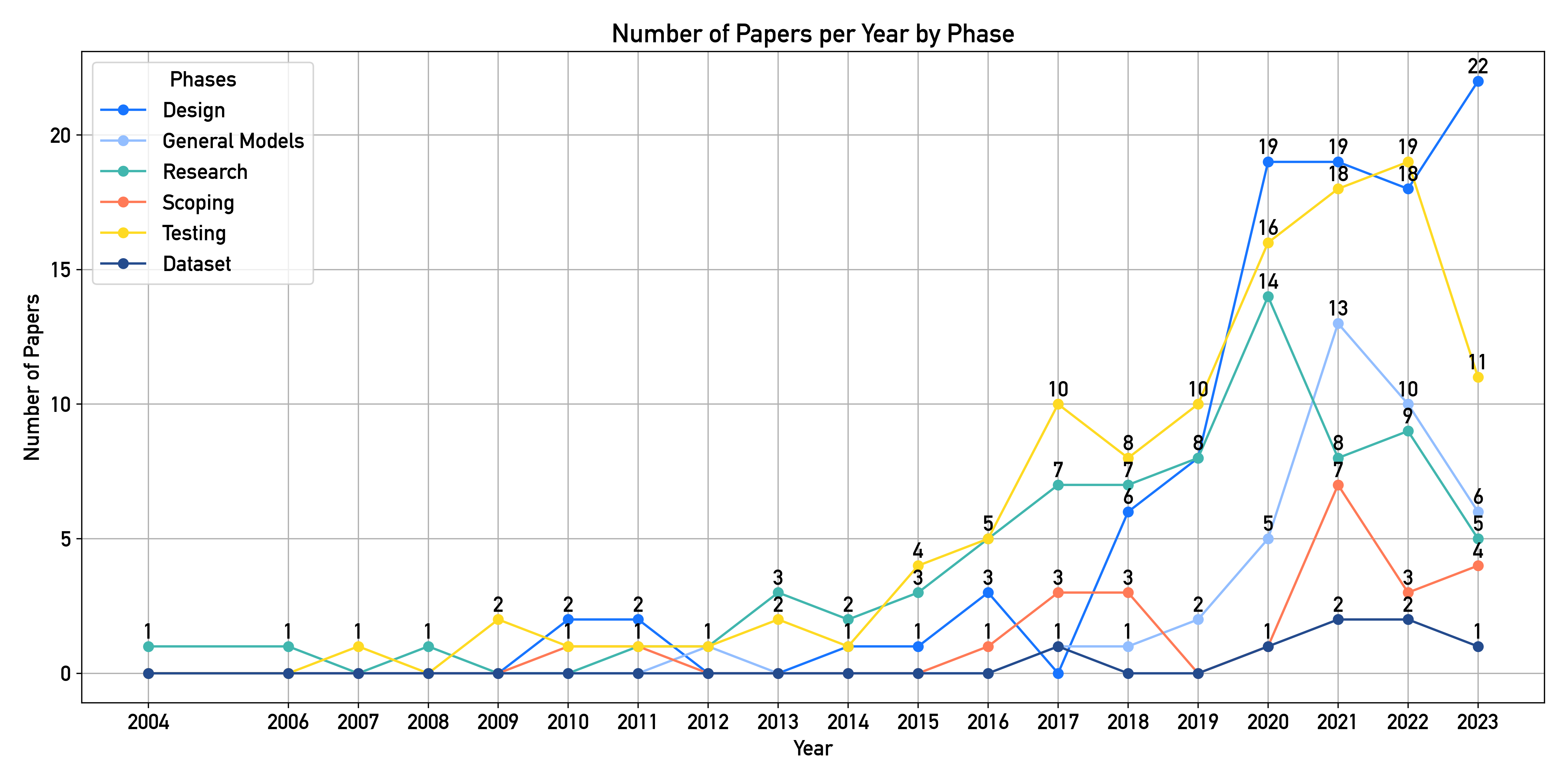}
    \caption{Trend of the number of papers each year in our sample, for each phase.}
    \label{fig:paper_year_phase}
\end{figure*}

\begin{figure*}[htbp]
    \centering
    \includegraphics[width=\textwidth]{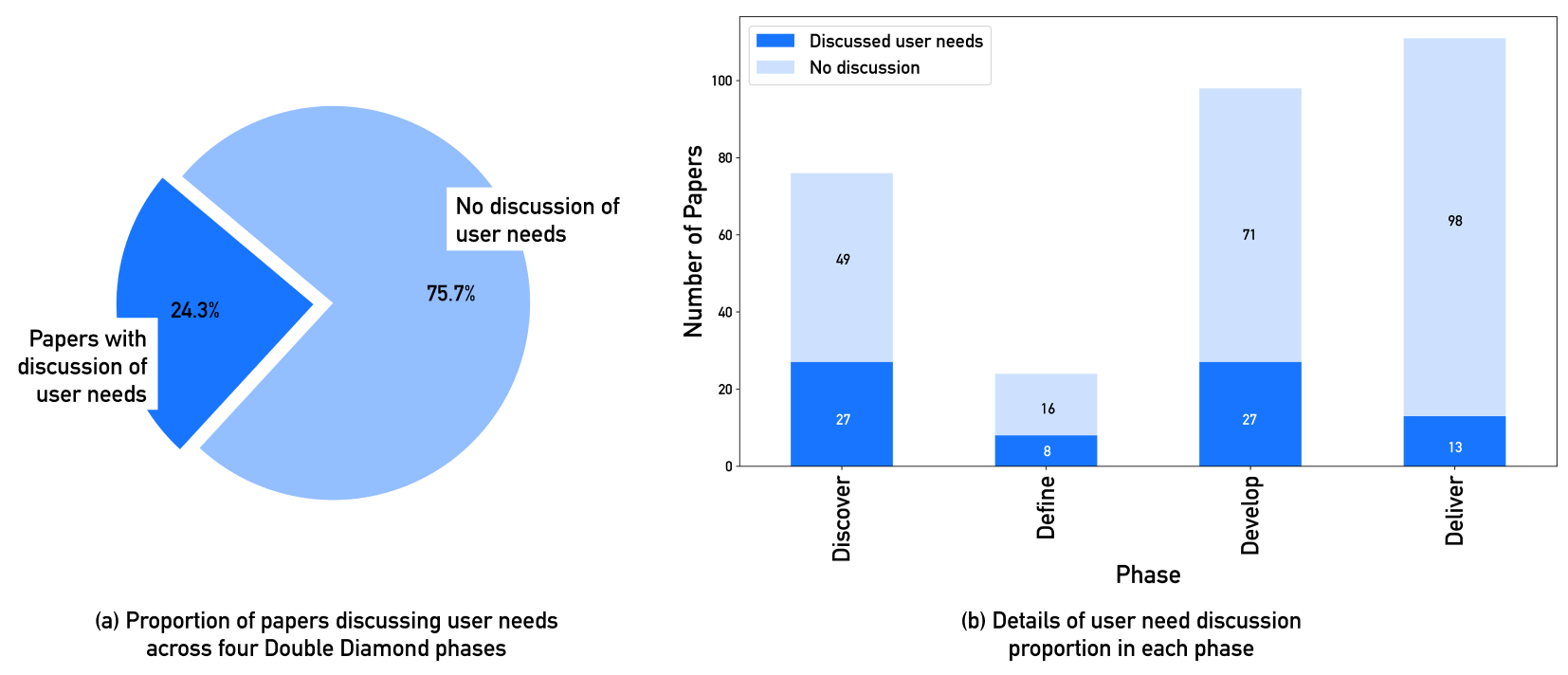}
    \caption{Classification of papers based on their consideration of user needs in each phase of the Double Diamond process. Our classification is based on whether they employed human-centered research methodologies, introduced in previous books and articles~\cite{olson_ways_2014, rosala_discovery_2020, farrell_ux_2017, rosala_how_2022, moran_quantitative_2018}, to understand their target user groups.}
    \label{fig:user_need_proportion}
\end{figure*}

Fig. \ref{fig:paper_year} depicts the trend of paper counts for each year in our sample, and Fig. \ref{fig:paper_year_phase} provides a more detailed view of the six phases. They show that research in this field has significantly increased since 2020. Note that the literature review was conducted from March to July 2023, so we only included papers published before this. Through further analysis of the general trends, we identified two imbalances in the current research landscape:

\paragraph{Imbalance between technology-centric and human-centered approaches}
We visualized the proportion of papers that studied or analyzed the needs of their target users using human-centered methodologies defined in previous literature~\cite{olson_ways_2014, rosala_discovery_2020, farrell_ux_2017, rosala_how_2022, moran_quantitative_2018}, such as ethnographic interviews, usability studies, etc. The result is shown in Figure \ref{fig:user_need_proportion}: in total, only 24.3\% papers (N=76) from all papers in these 4 phases (N=309) used human-centered methodologies and discussed user needs in their scenarios. This reflects the current \textit{technology-centric} tendency of research in AI assistance for UX. Although this phenomenon is not uncommon given the nascent nature of this field, it also calls for a more balanced approach that incorporates human-centered investigations. Emphasizing human-centered research not only addresses the preferences of users but also enhances the overall value and impact of AI solutions~\cite{shneiderman_human-centered_2022}.

\paragraph{Imbalance between studies in Double Diamond phases}
As depicted in Fig. \ref{fig:user_need_proportion} (b), the papers in our sample display a noticeable inclination towards the \textit{develop} and \textit{deliver} phases, while seemingly underrepresenting the \textit{define} phase. Determining the exact cause of this observed trend is challenging and beyond the scope of our review. Nevertheless, we hypothesize that this bias stems from the wealth of data available for the latter two phases (as discussed in Section \nameref{section:datasets}), coupled with the inherently subjective and task-dependent nature of evaluating design concepts during the \textit{define} phase~\cite{british_design_council_double_nodate, gray_its_2016}.

In the following sections, we dive deeper into our analysis of previous work in each of these six categories. At the end, we compare our findings from all six phases and summarize the results of our general analysis.

%% file: sections/4-1-discover.tex
\subsection{Discover}
\label{section:discover}
Discover is the divergent phase in the first diamond. It is the beginning phase where most exploratory user research is conducted. Designers need to understand the design problems and build user empathy in this phase. Common methodologies and artifacts involved in this phase include personas, user interviews, and brainstorming~\cite{british_design_council_double_nodate}. Our analysis summarized related research themes from past works as follows: \textit{Review Mining} (N=27), \textit{Data-driven Persona} (N=21), and \textit{AI-supported Brainstorming} (N=18).

\subsubsection{Review Mining}
For UX researchers, analyzing user reviews helps them identify current design problems, potential user requirements, and other user experience-relevant information~\cite{hedegaard_extracting_2013, baj-rogowska_exploring_2023, yang_exploiting_2019, mendes_uux-posts_2017}. The old-fashioned practice is manually coding data or using rule-based algorithms to classify user reviews into several topics and conducting statistical analysis~\cite{meiselwitz_investigating_2015, maalej_automatic_2016}. The introduction of machine learning to this task could date back to the 2010s~\cite{dabrowski_analysing_2022}. It automated the process of addressing vast amounts of textual data and advanced traditional algorithms with a better understanding of natural language. For example, it facilitates extracting structured information from the narratives, such as product features and user attitudes ~\cite{tuch_analyzing_2013}. 

One concern is how these works quantify the goal of mining user reviews. Most of them simplify design practitioners’ needs to classify user narratives based on some empirically defined computational models~\cite{hedegaard_extracting_2013, yang_exploiting_2019} or quantified metrics such as user sentiments~\cite{guzman_how_2014, li_extraction_2020}, satisfaction levels~\cite{jang_satisfied_2022, jang_modeling_2017}. Only a limited number of these works validated the effectiveness of this equivalence in meeting designers’ needs. 

Another concern is the generalizability of these formulations in identifying design problems in different scenarios. Some recent works indicated that review analysis could be fine-grained to user needs by integrating more advanced language models. For example, Wang et al.~\cite{wang_where_2022} increased the granularity of the extracted information and indicated specific problematic features for further improvements.

\subsubsection{Data-Driven Persona} 
Data-driven persona refers to the adoption of algorithm methods to develop personas from numerical data~\cite{salminen_survey_2021}. Machine learning pushes the purview further with its capacity in clustering and segmenting a variety of user data, such as feedback posts~\cite{tan_generating_2022, zhang_data-driven_2016, jisun_automatic_2017} and survey responses~\cite{degen_method_2020}. Besides, it also makes large-scale user data with time-changing behaviors feasible for persona development. For instance, user profiles and their interaction history~\cite{jansen_creating_2019, salminen_survey_2021, an_towards_2016} are introduced and make the persona construction more comprehensive. 

A common criticism of this data-driven approach is its automation process hinders design practitioners from building as deep user empathy as they could with a qualitative approach ~\cite{salminen_literature_2020}. Efforts have been made to integrate mixed methods in recent years. For instance, quantitative results are considered archetypes and inform the following qualitative analysis~\cite{tan_generating_2022, zhang_data-driven_2016, jansen_creating_2019}. Some other approaches verified qualitative insights via quantitative results~\cite{jung_two-handed_2022}. However, evaluations of these mixed-method approaches are far from standardized and overlook examining their effectiveness as user-empathizing processes. 

\subsubsection{AI-supported Brainstorming}
Ideation is another divergent thinking scenario with which many studies have tried to integrate AI. Research on AI for brainstorming includes individual support and human-human collaboration support.

For individual ideation supports, early systems adopted machine learning for retrieving inspirational ideas and searching associative knowledge from a defined collection~\cite{gilon_analogy_2018, feng_gallery_2022, andolina_inspirationwall_2015, kita_v8_2018}, among which only a limited number of works considered learning from specific design contexts~\cite{koch_may_2019}. Recently, the advancement of Large Language Models (LLMs) has enhanced the capacity for divergent thinking in these ideation systems, but it also confines them primarily to textual modalities~\cite{memmert_towards_2023, lopez_triggering_nodate, di_fede_idea_2022}. Besides, these systems mostly followed a series of linear structured stages in order to make AI integration more feasible~, like a sequence model consisting of warm-up, generating ideas, discussing ideas with groups ~\cite{lopez_triggering_nodate,memmert_towards_2023, zaphiris_conversational_2020}.
 
For collaborative brainstorming, researchers investigated how machine learning could be involved and support various interactions for team communication, like face-to-face ideation~\cite{andolina_inspirationwall_2015} and table-top interfaces~\cite{hunter_wordplay_nodate}. Machine learning could also be a team facilitator for human group ideations~\cite{bittner_designing_2019, zaphiris_conversational_2020}. The prosperity of generative AI provides more engaging roles for machine learning~\cite{shin_integrating_2023} such as experts~\cite{memmert_towards_2023, bittner_designing_2019} and mediators~\cite{lobbers_ai_2023}, and leads to further research opportunities.

Another research focus is on how social effects in human-human teaming transfer in human-AI collaborative ideation~\cite{hwang_ideabot_2021, memmert_towards_2023}, which sheds light on the negative impacts that AI has introduced in this process, like distraction~\cite{kita_v8_2018}, cognitive loads~\cite{zhang_melting_2022}, and free-riding~\cite{memmert_towards_2023}, which are not limited to UX ideation.

\subsubsection{Additional Topics}

In addition to the aforementioned topics, some other emerging works merit mention. Some studies addressed challenges in traditional qualitative research, such as communication fatigue and evaluation apprehension, by introducing AI-powered conversational agents~\cite{xiao_tell_2020, bulygin_how_2022}. Researchers have explored its adoption in conducting user interviews, facilitating engaging communication with users, and information elicitation~\cite{han_designing_2021, xiao_if_2020}. Moreover, conducting user interviews at scale would be more accessible. How this interview mode affects interviewers, interviewees, and in-depth understanding leaves opportunities for future studies.

\subsubsection{Summary}

Current ML integrations in UX research mostly provide automation support for laborious work and enhance traditional processes in work with large-scale and large-variety user data, especially for review mining and data-driven persona. Studies on ideation are more diverse and consider different collaborative settings and potential roles of AI beyond automation. Based on a human-centered scope, an apparent question is how the integration of machine learning aligns with the need of the UX discover phase, which is, understanding design problems and building user empathy. 

What we found from our analysis is \textbf{an oversight of the empathy-building process} and \textbf{a limited interpretation of design practitioners’ needs}. For example, constructing personas is regarded as making deliverables that could be automated by machines, while it is primarily a process where designers synthesize materials and build user understanding; quantitative metrics are adopted without validating its effectiveness for design practitioners and its generalizability for various design contexts. Future studies would be enriched by delving deeper into specific design contexts and designers' cognitive processes, especially in enhancing the empathetic comprehension of users as highlighted by~\cite{zhu_toward_2023}. This should complement the focus on the informational necessities that bolster the designers' empathetic processes.

%% file: sections/4-2-define.tex
\subsection{Define}
\label{section:define-phase}
\textit{Define} is the convergent phase in the first diamond, where designers \textbf{define the problem statement} and \textbf{pinpoint the products' desired impact} based on previous research findings. The main themes we identified in the \textit{define} phase are \textit{Qualitative Analysis} (N=22) and AI for \textit{Design Idea Evaluation} (N=2)\footnote{We specifically looked for other papers involving design idea evaluation and AI, but did not find any beyond our snowball sampling results.}. Methodologies and artifacts involved in this phase include affinity diagramming and focus groups~\cite{british_design_council_double_nodate}. The primary objective of this phase is to sort through the extensive research data, discerning the most promising directions that align with user requirements, business objectives, and technical viability~\cite{rosala_how_2022}.

\subsubsection{Qualitative Analysis}
\label{section:qualitative_analysis}
AI support for qualitative analysis has been an active research area and is prevalent in our sample (N=22). UX professionals and HCI researchers use this methodology to organize, label, and analyze data, to identify patterns and extract insights~\cite{rosala_how_2022, olson_ways_2014}. Generally, researchers discovered that simplistic automation of qualitative analysis can break established workflows, increase discussion overhead, and lead to unexpected reductions in efficiency and quality  ~\cite{borlinghaus_comparing_2021}. In contrast, papers that closely examined different steps in qualitative analysis and intentionally preserved human agency, control, and goals often demonstrated better psychological and performative results~\cite{marathe_semi-automated_2018, rietz_cody_2021, wakatsuki_clustering_2021, feuston_putting_2021, gebreegziabher_patat_2023, gao_collabcoder_2023}
    
    On the surface, qualitative analysis involves \textbf{labeling data} and \textbf{extracting insights}. Some studies aimed at speeding up the labeling process and using AI to produce labeled results~\cite{li_qualitative_2021}. However, research has shown that these \textit{full} automation approaches can easily break the existing workflow and lead to increased discussion overhead and reduced efficiency and quality~\cite{borlinghaus_comparing_2021}. In contrast, some papers broke down detailed steps in qualitative analysis to analyze their distinctions and different potentials for automation. Marathe et al.~\cite{marathe_semi-automated_2018} divided qualitative analysis into two phases: \textbf{building a codebook} by analyzing data, and \textbf{applying the codes} to the remaining data. 
    
    \paragraph{Codebook building} Building a codebook with a data subset is a key learning and reasoning process in qualitative analysis, where researchers build \textit{``emotional connection --- the intimacy, pride, and ownership --- with the data''}~\cite{jiang_supporting_2021} and \textit{``think with their hands''}~\cite{borlinghaus_comparing_2021}. Researchers generally \textbf{oppose the introduction of \textit{``low-level, suggestion-based automation''}} in this process, to avoid taking away the invaluable cognitive process of human researchers~\cite{marathe_semi-automated_2018, jiang_supporting_2021}. Feuston et al.~\cite{feuston_putting_2021} emphasized that qualitative research is a process that utilizes researchers' unique perspectives in data analysis, whilst AI might take away this opportunity and reinforce past coding patterns in new data.

    \paragraph{Codebook application} Once a codebook is developed, applying it to the remaining data can be relatively more mechanical. Previous studies have shown that \textbf{automation is more welcomed in this phase}~\cite{marathe_semi-automated_2018}. As a result, many systems were built to automate the tedious aspects of labeling while preserving the researcher's agency in learning~\cite{marathe_semi-automated_2018, rietz_cody_2021, gebreegziabher_patat_2023, jiang_supporting_2021, feuston_putting_2021}. But there is also more to the labeling process than simply applying the codebook: since qualitative analysis is often a collaborative process, Drouhard et al. emphasized the value of disagreement between researchers in reflecting ambiguities in data~\cite{drouhard_aeonium_2017}. Reflecting on and resolving these conflicts can help to \textit{improve} researchers' learnings~\cite{chen_using_2018, rietz_cody_2021, gebreegziabher_patat_2023}. 

    \paragraph{Advantages of interactive ML in qualitative analysis} The interactive ML technique provides great potential to automate tedious aspects of qualitative coding, while leaving the final decisions to users, preserving their agency. It has been employed in existing AI systems to support qualitative coding~\cite{rietz_towards_2020, rietz_cody_2021, gebreegziabher_patat_2023}. In the context of qualitative analysis, interactive ML engages users in a collaborative process, where they actively offer feedback on AI-generated outputs, thereby enhancing the precision and relevance of qualitative coding~\cite{rietz_towards_2020}. Interactive ML also does not require large labeled datasets and learns as users annotate more data, which naturally fits the qualitative analysis process. Building on top of human-interpretable rules, patterns and relatively simple AI models, they were able to achieve a certain level of explainability and interpretability. \textsc{Cody} also provided counterfactual explanations to help users further understand algorithmic predictions.

    \paragraph{User control in qualitative analysis} User control has been a common theme in discussions of AI support in qualitative analysis~\cite{ rietz_towards_2020, jiang_supporting_2021, feuston_putting_2021, rietz_cody_2021, gebreegziabher_patat_2023, gao_collabcoder_2023}. Earlier papers discussed how the lack of control might prevent AI from providing valuable support~\cite{jiang_supporting_2021}. However, Feuston and Brubaker discovered that it is more nuanced: AI support can benefit certain steps in qualitative analysis, or even shifting some analytic practices, as long as it assists instead of automates existing analytic work practices~\cite{feuston_putting_2021}. The careful design of systems including \textsc{Cody}~\cite{rietz_cody_2021} and \textsc{PaTAT}~\cite{gebreegziabher_patat_2023} also confirmed the value of AI support while maintaining user control and agency. The ``delegability'' of human tasks to AI~\cite{lubars_ask_2019} in qualitative coding depends on human motivation, task difficulty, associated risk, and human trust~\cite{jiang_supporting_2021}.


    \subsubsection{Design Idea Evaluation}
    \label{section:design-idea-evaluation}
    Two papers in our sample investigated the use of AI in evaluating design ideas. Siemon conducted a comparative study with a simulated AI system to investigate AI's utility in helping reduce apprehension in design idea evaluation~\cite{siemon_let_2023}. In addition, Mesbah et al. combined AI with crowdsourcing to effectively measure the desirability, feasibility, viability, and overall feeling of design ideas~\cite{mesbah_hybrideval_2023}. Given that current methodologies around design idea evaluation are subjective and task-dependent~\cite{british_design_council_double_nodate, gray_its_2016}, AI models that are trained against general metrics such as in~\cite{mesbah_hybrideval_2023} are likely not sufficient for real-world scenarios. It remains largely unclear how AI support might fit into existing manual evaluation processes. We believe a deeper empirical understanding of UX evaluation processes and practices is required to bridge this current gap.

\subsubsection{Summary}
    
    In all, in the \textit{define} phase, previous research that emphasized researchers' agency in understanding, learning, and interpreting data with their unique perspectives generally showed better results than simplistic automation and acceleration~\cite{marathe_semi-automated_2018, rietz_cody_2021, wakatsuki_clustering_2021, feuston_putting_2021, gebreegziabher_patat_2023, gao_collabcoder_2023}. The use of interactive ML techniques in qualitative analysis support has demonstrated potential in balancing researchers' agency in learning and interpreting the data with algorithmic support~\cite{rietz_cody_2021, gebreegziabher_patat_2023}. For evaluating design ideas with AI, the subjective and task-dependent nature of current evaluation practices~\cite{british_design_council_double_nodate, gray_its_2016} requires closer coupling between designers' workflows, goals, and AI support to provide meaningful, holistic support.

    


%% file: sections/4-3-develop.tex
\subsection{Develop}
\label{section:develop}
Develop refers to the divergent phase where designers come up with solutions for the defined problem domain, informed by insights from the previous two phases~\cite{british_design_council_double_nodate}. Our analysis identified the following themes for papers in this phase: \textit{UI Generation} (N=51), \textit{Interface Design Inspiration} (N=25), \textit{UI Optimization} (N=21).

\subsubsection{UI Generation}
Large-scale UI datasets like RICO enabled AI research in automatic UI generation (more discussion about datasets is in the \nameref{section:datasets} section). We divide past UI generation research roughly into 3 categories: full-screen UIs, UI components, and fidelity conversion. 

\paragraph{Full-screen UIs}
Many previous AI models focused on generating entire UI screens. As a fundamental step to effectively automate structuring UI elements, layout generation becomes a predominant focus of many previous work. 
Earlier on, Li et al. proposed applying  Generative
Adversarial Networks (GANs) to synthesize and model geometric relations of graphical elements for accurate layout alignment~\cite{li_layoutgan_2021}.
Furthermore, transformer-based architectures~\cite{gupta_layouttransformer_2021, jiang_layoutformer_2023, sobolevsky_guilget_2023} provided solutions that handle the hierarchical and sequential relationships of graphical elements, adding value especially for the UI generation task. Along the same line, Inoue et al.~\cite{inoue_layoutdm_2023} and Zhang et al.~\cite{zhang_layoutdiffusion_2023} leveraged diffusion models for conditional layout generation. 
While these efforts mark considerable progress, the generation of high-fidelity UI screens remains early-stage, with notable attempts such as GUIGAN by Zhao et al.~\cite{zhao_guigan_2021}, approaching high-fidelity generation through integrating GUI component subtree sequences in the generation process.
Overall, we found only few existing AI models that offer high-fidelity UI generation ready for use in practice. The trajectory of UI layout and high-fidelity UI generation research reveals the critical need for solutions that are directly applicable in design workflows. Despite the trend towards more sophisticated AI capabilities, there remain unresolved challenges and gaps to seamlessly blend model-generated results with user-centered design practices. 

\paragraph{UI Components}
A few papers were dedicated to the generation of UI components, such as icons~\cite{zhao_iconate_2020} and buttons. For example, ButtonTips ~\cite{liu_buttontips_2019} dived deeply into automatic web button design with user input constraints, including button layout generation with text labels, color selection, spatial relationships, and presence prediction. These research efforts can help generate need-based design resources for novice designers. Additionally, designers in the industry nowadays commonly work with company-specific design systems to ensure branding and visual consistency~\cite{frost_atomic_2016}. Generation within the constraints of design systems might increase the adoption of AI tools in design practitioners' workflow.

\paragraph{Fidelity Conversion}
Except for AI models that adopt an end-to-end approach for UI generation, past research also investigated AI models' capabilities in converting UI prototypes between different fidelities~\cite{buschek_paper2wire_2020}. For example, Paper2Wire turns UI sketches into editable, mid-fidelity UI wireframes~\cite{buschek_paper2wire_2020}, which can be helpful for early prototyping stages. MetaMorph, for another instance, assists in transforming constituent components from lo-fi sketches to higher fidelities~\cite{sermuga_pandian_metamorph_2021}. Rather than directly delivering the final result, such AI models take an apporach to facilitate designers' existing workflows and contain a higher potential for adoption.

\subsubsection{Interface Design Inspiration}
Designers usually refer to external resources for inspiration. Currently, prevalent applications of example search fall into two categories: (1) design galleries, such as \emph{Gallery D.C.}~\cite{feng_gallery_2022}, where designers usually browse a wide range of examples as a serendipitous inspirational process; (2) algorithmic recommendation tools~\cite{swearngin_rewire_2018} based on similarities to the user's design input, where designers look for suggestions focusing on more concrete ideas~\cite{mozaffari_ganspiration_2022}. Previous studies showed two challenges of existing exploratory strategies: design fixation (e.g. excessive focus on present concern)~\cite{marsh_how_1996, youmans_design_2014} and focus drift (e.g. deviation from original goal). Intelligent tools such as GANSpiration~\cite{mozaffari_ganspiration_2022} generate diverse but relevant design examples, which seek the balance between and provide both targeted and serendipitous inspiration. Scout, for another example, focused on overcoming design fixation, providing more spatially diverse design examples, and ``breaking out the linear design process``~\cite{swearngin_scout_2020}. Meanwhile, AI might shed light on scaling up earlier solutions that help to avoid design fixation, such as parallel prototyping, by supporting exploring relevant alternatives during iteration~\cite{dow_parallel_2011}.

Example exploration usually takes place in the early stages of design and continues to be a crucial component throughout the iterative process, expanding potential solution space. Existing AI-infused tools for inspiration search have expanded diversity of search mediums, enabling inputs such as such as natural language description~\cite{wang_screen2words_2021}, screenshots~\cite{swearngin_rewire_2018}, hand-drawn sketches and doodles~\cite{mohian_psdoodle_2022}, low-fidelity design artifacts such as wireframes~\cite{chen_wireframe-based_2020}, and hybrid inputs (e.g. text and doodle~\cite{mohian_searching_2023}), supporting more flexible search processes~\cite{lu_bridging_2022}. In later stages of design, external references also allow for reinterpretation of ideas and are used as validation tools~\cite{herring_getting_2009}. 
Given the iterative nature of design tasks, more research is needed on dynamically supporting and inspiring UI design as the artifact evolves in complexity and fidelity. 

\subsubsection{UI Optimization}

UI optimization encompasses two main aspects:
at the interface level, it involves enhancing the layout positioning and aesthetic style ~\cite{rahman_ruite_2021}; at the user experience level, it focuses on improving the perceived affordances of components~\cite{swearngin_modeling_2019, pang_directing_2016}. The process mainly aim at optimizing visual appeal, functional clarity, are addressed, and the overall interaction with the user interface.
First, applying appropriate visual aesthetics plays an important role in generating and optimizing high-fidelity UI. The underlying difficulties in automatically suggesting and applying design styles include data-driven aesthetic assessment~\cite{kong_aesthetics_2023, kumar_dynamic_2023} and transforming high-level design principles into explicit constraints. Accordingly, researchers proposed solutions that 1) translate natural language requirements into predictions of design properties~\cite{kim_stylette_2022} and 2) extract applicable design constraints from design principles~\cite{kong_aesthetics_2023}. There are also a few papers dedicated to specific aspects of aesthetics, such as color~\cite{feng_smartcolor_2021, hegemann_cocolor_2023, odonovan_color_2011} and font design~\cite{zhao_modeling_2018, odonovan_exploratory_2014}. Meanwhile, due to the subjectivity of aesthetic styling, existing systems tend to keep designers actively engaged in the producing process, including making decisions about which recommended suggestions to adopt, iterating on their choices, and making further revisions afterwards~\cite{kong_aesthetics_2023, kim_stylette_2022, hegemann_cocolor_2023}.
For optimization at the user experience level, past work drew insights from the correlation between components' spatial relationships and user task performance (i.e. speed and accuracy), leveraging classic principles such as Fitts's Law and neural network learning~\cite{duan_optimizing_2020} to reach ideal layout. Different from the previous categories, optimization contributes to finishing the design cycle. Given the standardized and consistent requirements across UI design practices, optimization tasks can further explore topics including visual alignment and consistency checking, usability issue mitigation, and design guidelines adherence improvement.

\subsubsection{Summary}
Machine learning, by enhancing design processes with its search and generative capabilities, offers innovative pathways for design inspiration~\cite{feng_gallery_2022}. AI-enabled search and generation might enable more rapid and parallel prototyping, previously limited by human capacity, thereby increasing the potential to elevate design outcomes. While the quest for end-to-end solutions for complete UI design remains prevalent, there's a shift towards automating select intermediary steps in the design workflow, promising more effective support for design objectives ~\cite{lu_bridging_2022}. Additionally, for design aspects steeped in subjectivity, like aesthetic choices, machine learning-assisted tools are emerging to bolster designers' creative freedom through detailed interactions, ensuring technology complements rather than overrides human expertise.

%% file: sections/4-4-deliver.tex
\subsection{Deliver}
\label{section:deliver}
Deliver is the convergent phase in the second diamond, where through different evaluation methods, designers elicit feedback from users on their design prototypes, iteratively improve them, and come up with a final solution~\cite{british_design_council_double_nodate}. There are several major themes in the testing phase: \textit{Visual Saliency Prediction} (N=24), \textit{Aesthetic Analysis} (N=12), \textit{Visual Error Detection} (N=9). 

\subsubsection{Visual Saliency Prediction}
\label{sec:saliency-prediction}
Visual saliency is a proxy of the perceived importance of screen components, indicating UIs' visual hierarchy. Such information can help UX practitioners better grasp users' attention distribution, thus improving the information architecture design~\cite{novak_eye_2023}.
Many model architectures have been developed for predicting visual saliency~\cite{xu_spatio-temporal_2016, georges_ux_2016, li_webpage_2016, bylinskii_learning_2017, shen_predicting_2015}. Visual attention prediction for different user groups ~\cite{leiva_modeling_2022, chen_learning_2023}, and UI categories~\cite{fosco_predicting_2020} allows more granularity and versatility for UX practitioners. Techniques to collect user gazing data with easy-to-access gadgets instead of expensive eye-tracking devices, such as webcams~\cite{xu_turkergaze_2015} and mobile phones~\cite{li_towards_2017}, have also been investigated. Methods deploying crowd-sourcing for data collection are also presented, with eye-tracking techniques ~\cite{xu_turkergaze_2015} and by self-reporting where they had gazed at~\cite{cheng_visual_2023}. 

\subsubsection{Aesthetic Analysis}
\label{sec:aesthetic-analysis}
Automatic visual aesthetic analysis of UI screens can help UX professionals grasp perceptions of their design. While judging the visual appearance of UIs can be subjective, automatic evaluations afford quick predictions as initial feedback to designers. Past work has focused on AI applications in the evaluation of UI's perceived aesthetics~\cite{lima_assessing_2022, miniukovich_computation_2015, de_souza_lima_automated_2022, xing_computational_2021, dou_webthetics_2019} and visual complexity~\cite{akca_comprehensive_2021}, which is a key aspect of design aesthetics.
In addition, aesthetic predictions according to different user groups~\cite{leiva_modeling_2022} and in real usage contexts~\cite{samele_bootstrapped_2023} facilitate more nuanced prediction needs. The majority of existing visual analyses of UIs relied on objective metrics and feature extraction~\cite{akca_comprehensive_2021}, or AI models trained on user ratings~\cite{dou_webthetics_2019, leiva_describing_2022}. Both empirical analysis and experiment results have demonstrated the improved flexibility and quality of AI models' evaluations ~\cite{akca_comprehensive_2021, dou_webthetics_2019}.

A study conducted by Rozenholtz et al. revealed that in practice, the perceived visual quality is not the only factor contributing to the evaluation of a design~\cite{rosenholtz_predictions_2011}. Designers often have to make trade-offs between visual quality and design goals, which they concluded, \textit{``would likely interfere with acceptance of a perceptual tool by professional designers''}. In addition, they observed that \textit{the overall ``goodness'' values were not useful} beyond A/B comparisons between design options. A deeper empirical understanding of how UX practitioners utilize UI evaluation tools in real-world contexts would greatly benefit practical research in this direction.

\subsubsection{Visual Error Detection}

Automated visual error detection for UI screens is another key theme. Those systems can emulate human interactions with UI screens and save time and human effort after app development~\cite{peng_mubot_2022}. While these systems are often used after app development and to check implementation quality, they are also capable of identifying design issues that get propagated to code implementation. Unlike system-specific tests like those developed especially for Android~\cite{collins_deep_2021, llacer_giner_deep_2020}, image-based testing techniques can take UI screenshots from different systems, increasing cross-platform versatility~\cite{eskonen_automating_2020,eskonen_deep_2019}. These automated testing techniques help detect display issues~\cite{su_owleyes-online_2021}, generate testing reports, and detect UI discrepancies between its design and development~\cite{chen_ui_2017}. Some specific techniques, such as interaction and tappability prediction~\cite{swearngin_modeling_2019, schoop_predicting_2022} can also be utilized to serve more granular error detection goals. Design guideline violation checkers~\cite{zhao_seenomaly_2020, yang_dont_2021,yang_uis-hunter_2021} also have great practical potential in UX workflows. Overall, AI has great potential in flexible and universal visual error detection.

\subsubsection{Additional Topics}

Systems around sentiment prediction, usability testing, and automatic feedback generation are also included in our sample. Sentiment prediction centers the user's perception of the product~\cite{desolda_detecting_2021, petersen_using_2020}. Some related works are user satisfaction prediction~\cite{koonsanit_predicting_2021, koonsanit_using_2022}, and brand personality prediction~\cite{wu_understanding_2019}. These models help guide the designer to analyze the design target and the predicted user perception. 

Usability testing is also one processes that gather the researcher's attention. To suit more nuanced device-specific usability testing needs, researchers present usability testing for mobile UI~\cite{schoop_predicting_2022}, e-learning system ~\cite{oztekin_machine_2013}, and thermostat~\cite{ponce_deep_2018}. Researchers use live emotion logs ~\cite{filho_automated_2015}, think-aloud sessions~\cite{fan_automatic_2020, fan2022human}, and online reviews~\cite{hedegaard_mining_2014} to extract usability-related data and assess interfaces. In addition, automatic feedback generation empowers designers to improve on the current design with the help of the ML system~\cite{krause_critique_2017, ruiz_feedback_2022}.

Other less-explored themes include dark pattern detection~\cite{hasan_mansur_aidui_2023} and A/B testing~\cite{kaukanen_evaluating_2020, kharitonov_learning_2017}. As accessibility design becomes more essential in UX design, researchers developed tools around automated accessibility testing ~\cite{vontell_bility_2019}. 


\paragraph{Summary} Previous research in the deliver phase has explored various ways to provide UI evaluation feedback to designers. We observed that in our sample, these explorations are often based on the visual analysis of UIs. However, with the growing prevalence of design systems in practice~\cite{frost_atomic_2016, churchill2019scaling}, UX designers are shifting their focus from pixel-level asthetics to interaction flows and the holisitc user experiences across UI screens. The evaluation of interaction flows and user experiences go beyond saliency prediction (Section~\ref{sec:saliency-prediction}) and visual asthetics (Section~\ref{sec:aesthetic-analysis}), yet is still overlooked in research. Moreover, current visual analysis metrics does not often align with unique UI design asthetics such as flat design and skeumorphism, restricting their practical adoption. We believe more considerations of these unique aspects of UX design is important in creating translational research value~\cite{colusso2017translational, colusso2019translational, norman2010research}

\begin{figure*}[htbp]
    \centering
    \includegraphics[width=0.9\textwidth]{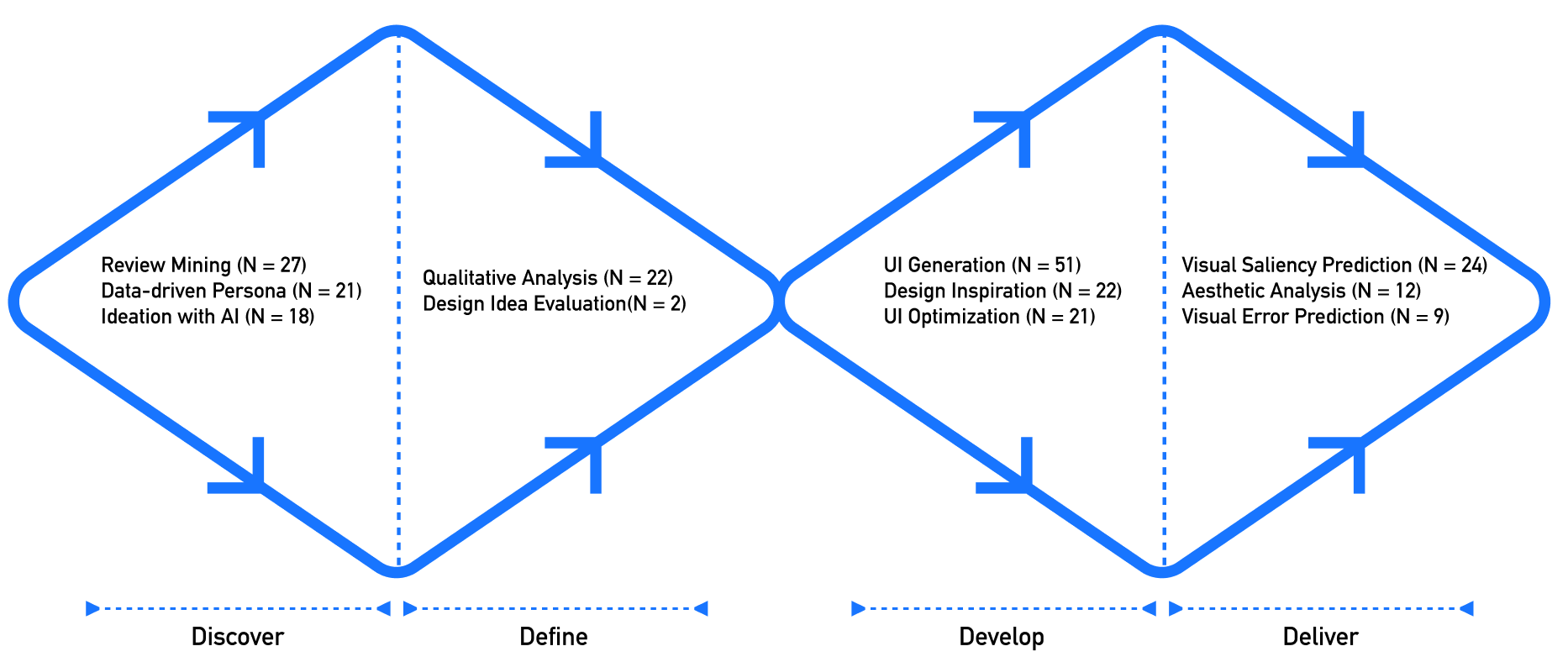}
    \caption{Double diamond with phase topic}
    \label{fig:double_diamon_2}
\end{figure*}

%% file: sections/4-5-datasets.tex
\subsection{Datasets}
\label{section:datasets}

\begin{table*}[!htbp]
\small
\renewcommand{\arraystretch}{1.4}
\begin{tabularx}{\textwidth}{L{0.125\textwidth}L{0.15\textwidth}XL{0.3\textwidth}L{0.25\textwidth}}
\toprule
\textbf{Category} & \textbf{Dataset} & \textbf{Year} & \textbf{Descriptions} & \textbf{Size} \\ \midrule

\multirow{5}{*}{\begin{tabular}[x]{@{}l@{}}Mobile Interfaces\end{tabular}} & RICO~\cite{deka_rico_2017} & 2017 & a large repository of Android app designs & 72k screens from 9.7k apps, 3M components  \\
\cline{2-5}
& ReDraw~\cite{moran_machine_2020} & 2018 & UI screens with GUI metadata & 14k screens, 191k components  \\
\cline{2-5}
& Enrico~\cite{leiva_enrico_2020} & 2020 & human-annotated topic modeling of RICO subset & 1.5k screens, 20 topics \\
\cline{2-5}
& VINS~\cite{bunian_vins_2021} & 2021 & wireframes and annotations for sketches and high-fidelity UIs for Android and iOS & 11 components, 257 wireframes, 4.5k high-fi screens \\
\cline{2-5}
& Screen2Words~\cite{wang_screen2words_2021} & 2021 & screen summarizations based on RICO & 112k summaries for 22k screens \\
\cline{2-5}
& Clay~\cite{li_learning_2022} & 2022 & human-made annotations to denoise RICO & 60k screen layouts \\
\cline{2-5}
& Android in the Wild~\cite{rawles_android_2023} & 2023 & human demonstrations of mobile device interactions & 715k interactions, 30k instructions \\
\cline{2-5}
& Swire~\cite{huang_swire_2019} & 2019 & crowd-sourced, hand-drawn sketches based on RICO & 3.8k screens   \\
\cline{2-5}
& UISketch~\cite{sermuga_pandian_uisketch_2021} & 2021 & crowd-sourced, hand-drawn low-fi UI element sketches & 18k sketches of 21 UI elements   \\
\cline{2-5}
& Synz~\cite{sermuga_pandian_synz_2021} & 2021 & synthetic smartphone low-fi screen sketches, generated based on RICO and UISketch & 175k screens   \\
\cline{2-5}
& Lofi Sketch~\cite{sermuga_pandian_lofi_2022} & 2022 & crowd-sourced, hand-drawn smartphone low-fi screen sketches, generated by random allocation & 4.5k screen sketches, annotated with 21 UI element categories  \\
\hline

\multirow{5}{*}{\begin{tabular}[x]{@{}l@{}}Web Interfaces\end{tabular}} & Webzeitgeist~\cite{kumar_webzeitgeist_2013} & 2013 & a large repository of web interfaces & 100k screens, 100M components  \\
\cline{2-5}
& WebUI~\cite{wu_webui_2023} & 2018 & low-cost, large-scale repository of web interfaces & 400k screens   \\
\cline{2-5}
& Webshop~\cite{yao_webshop_2023} & 2022 & human demonstrations of e-commerce website interactions & 12k instructions, 1.6k demonstrations  \\
\bottomrule
\end{tabularx}
\caption{Overview of public datasets identified in literature review sample, categorized by interface type and detailed with descriptions, sizes, and publication years.}
\label{ui_datasets_table}
\end{table*}

Open-sourced datasets on user interfaces of different devices and modalities have significantly contributed to AI support for UI/UX design. In Table~\ref{ui_datasets_table}, we summarized open-sourced datasets we collected in our sample. We discovered that \textbf{current datasets often overlook the user experiences underlying the interfaces}, limiting their applications in UX design. As a result, technical work based on these datasets often leans toward the latter two phases of \textit{develop} and \textit{deliver} in the Double Diamond (Fig.~\ref{fig:double_diamond}). In addition, we still \textbf{lack universal benchmarks to evaluate design quality}, reflected in UIs and the underlying user experiences, due to the often diverse goals of design in different digital products. Overall, existing datasets have enabled more technical solutions to user task automation~\cite{li_sugilite_2017, rawles_android_2023} and design task automation~\cite{moran_machine_2020, arroyo_variational_2021, huang_creating_2021}, instead of designer-centric augmentation tools.

\subsubsection{Mobile UI Datasets}
Most of the publicly available datasets for UX-related tasks focus on mobile user interfaces. RICO~\cite{deka_rico_2017} is arguably the most utilized UI dataset, containing 72k mobile screens and 3M UI components. \textsc{ReDraw}~\cite{moran_machine_2020} is a similar mobile UI dataset with 14k screens and 191k annotated components. Later on, many papers aimed at augmenting RICO in different directions, including \textit{topic modeling}~\cite{leiva_enrico_2020}, \textit{semantic summarization}~\cite{wang_screen2words_2021}, \textit{element mismatch denoising}~\cite{li_learning_2022}, and \textit{adding new screens and wireframes}~\cite{bunian_vins_2021}. In addition, Android in the Wild (\textsc{AitW}) is a mobile dataset containing 715k interaction episodes, spanning 30k unique instructions on different Android devices~\cite{rawles_android_2023}.

Four sketch datasets have also been released for mobile UI screens and components. \textsc{Swire} was created by recruiting designers to hand-draw sketches of 3.8k interfaces taken from RICO~\cite{huang_swire_2019}. Similarly, UISketch crowdsourced 18k hand-drawn low-fidelity UI elements~\cite{sermuga_pandian_uisketch_2021} and Lofi Sketch crowdsourced 4.5k screen sketches~\cite{sermuga_pandian_lofi_2022}. The Synz dataset took a purely synthetic approach and generated UI screen sketches with UI elements in UISketch and UI layouts from screens in RICO~\cite{sermuga_pandian_synz_2021}.

\subsubsection{Web UI Datasets}
Given the significant resources and restrictions involved in collecting mobile UI data, researchers also collected website UI datasets: \textsc{Webzeitgeist}~\cite{kumar_webzeitgeist_2013} with 100k pages and 100M elements, and \textsc{WebUI}~\cite{wu_webui_2023} with 400k pages. The advantage of collecting web UI is the ability to scale up with responsive layouts in different viewport sizes. Many websites do not require logins to view content, avoiding the potential login wall in mobile app UI collections~\cite{wu_webui_2023}. Recent studies also demonstrated the potential to augment AI models' understanding of mobile UIs with web UI data~\cite{wu_webui_2023}. A UI navigation and automation dataset for e-commerce websites, \textsc{WebShop}, with 12k crowd-sourced text instructions and over 1.6k human demonstrations is also created~\cite{yao_webshop_2023}.

\subsubsection{Discussion}
\label{section:dataset-discussion}

\paragraph{Missing connections across UI screens} Most datasets in our sample consist of \textit{individually separated} UI screenshots, their hierarchy information, and metadata. They miss the connections \textit{across multiple UI screens}, which encapsulate the underlying user tasks, experiences, and goals. The absence of these inter-screen connections highlights a fundamental distinction between UI and UX design, limiting existing research's practical applications in the UX industry~\cite{norman1998definition}\footnote{The RICO dataset~\cite{deka_rico_2017} included interaction tracing and animation between screens, but they remain much underexplored, especially compared to the other parts of the dataset such as UI screenshots, hierarchies, and layouts.}. As a result, research based on these datasets tends to be biased towards viewing UIs predominantly as static, multimodal entities comprising textual and visual information~\cite{moran_machine_2020, li_screen2vec_2021, wang_screen2words_2021, huang_swire_2019, wang_enabling_2023}. This also led to a research focus skewed towards the \textit{develop} and \textit{deliver} stages of the Double Diamond model (Fig.~\ref{fig:double_diamond}), where static UIs appear more than the first two exploratory phases (Fig. \ref{fig:paper_year_phase}). We believe that a deeper understanding of UX practices and mindsets is essential to align datasets and AI models with the complexities of real-world UX design.

The recent release of datasets for UI task automation, such as Android in the Wild~\cite{rawles_android_2023}, provides valuable data on user flows across multiple UI screens. While the primary focus is on supporting task automation for end users, they also have the potential to benefit UX practitioners. For example, assessments of these user flows and their design contexts can help UX designers find relevant and high-quality inspirations in the early stages of design. In addition, commercially available, designer-centric datasets such as Mobbin\footnote{https://mobbin.com} can inform academic creation of open-source datasets that more directly afford applications in realistic UX domains.

\paragraph{Lack of meaningful evaluation metrics}
\label{section:discussion-2}
We still lack objective metrics that effectively reflect the quality of UI and the underlying user experiences. Most AI generation models trained on existing datasets are evaluated against metrics including overlap, alignment, and intersect-over-union (IoU) that do not necessarily align with the perceived quality of UIs~\cite{jing_layout_2023, li_layoutgan_2021, kikuchi_constrained_2021}. Other common metrics include visual complexities~\cite{alemerien2014guievaluator, riegler2015ui, reinecke2013predicting, ines2017evaluation}, visual saliency~\cite{zhao_guigan_2021, leiva_describing_2022, bylinskii_learning_2017, kumar_dynamic_2023, li_webpage_2016, shen_predicting_2015, kruthiventi_deepfix_2017, judd_learning_2009}, and visual similarities~\cite{li_screen2vec_2021, huang_swire_2019, karimi_creative_2020}.

This reflects the contrast between AI's data-driven nature and UX's user-centered philosophy, which contains opportunities for both disciplines and remains to be further explored~\cite{chromik_ml_2020}. Most UX tasks hardly be holistically evaluated using only objective metrics. In practice, they are often embedded in individual projects' contexts of user needs, business objectives, and technical feasibility. Evaluating UI screens and their user experiences significantly differs from traditional image-based assessment in domains like computer vision, requiring the development of novel, UX-focused objective metrics tailored for AI's application in this field.



%% file: sections/4-6-general-ai-models.tex
\subsection{General AI Models}
\label{section:general-ai-models}

In our sample, we also identified 37 AI-focused papers that do not specifically fit into any of the four Double Diamond phases, but still work with UX- and UI-related tasks. We analyze these papers here to further understand the current technical landscape. Generally, we identified three themes: (1) UI annotation \& component detection; (2) UI semantic understanding; (3) UI interaction automation. 
These foundational AI explorations from the 37 papers contain significant potential to assist UX designers through downstream tasks, providing practical applications that can enhance their design processes and outcomes.

\subsubsection{UI Annotation \& Component Detection}
Detecting and annotating visual elements on UIs can provide value in downstream tasks like UI semantic understanding and interaction automation, and also many different use cases including accessibility and UI testing~\cite{chen_towards_2022}. In our sample, researchers utilized many AI model architectures for this task, but ResNet~\cite{chen_towards_2022, li_widget_2020, chen_unblind_2020} and Faster-RCNN~\cite{zhang_screen_2021, manandhar_magic_2021} remain dominant given their impressive capabilities in general object detection. For component detection specifically on UIs, the precision for UI component locations and sizes is paramount, which slightly differs from general object detection. To address this challenge, some papers also included UI view hierarchies in addition to screenshot images for more accurate location information~\cite{zang_multimodal_2021, li_widget_2020}. Annotating UI components on screens can augment UI datasets with detailed meta-level information, supporting modular design paradigms such as Atomic Design~\cite{frost_atomic_2016}. \looseness=-1

\subsubsection{UI Semantic Understanding}
Many AI models were developed to tackle the fundamental task of UI understanding. They mostly focused on the semantic meaning, i.e., the functionalities and purposes, of UI components and screens. Many AI models take in screenshots and/or corresponding view hierarchies~\cite{li_screen2vec_2021, ang_learning_2022, li_vut_2021, bai_uibert_2021, wu_screen_2023} and output an embedding of the interface screen or component. ActionBert used user actions with the UI to learn a UI embedding~\cite{he_actionbert_2021}. Fu et al. made the analogy between words--sentences in NLP and pixels--screens for UI understanding. Recently, with the rise of increasingly larger model sizes, a relatively large vision-only UI model based on pre-trained large ViT and T5, Spotlight, was trained on 2.5M mobile screens and 80M web pages and achieved SoTA on some representative UI tasks~\cite{li_spotlight_2023}. These general-purpose models lay the groundwork for more sophisticated downstream tasks that can support various UX workflows. 

Another approach to achieving holistic UI understanding is through UI screen summarization. These summarizations present concise textual information regarding a UI screen's appearance and functionality, which can be useful for many language-based application scenarios. Researchers have attempted to use multimodal AI models~\cite{wang_screen2words_2021} and vision-based approaches~\cite{leiva_describing_2022} to generate such summaries. Such summaries can be helpful for text-based retrieval of similar screens, screen readers enhancement, and screen indexing for conversational applications~\cite{wang_screen2words_2021}.

\subsubsection{UI Interaction Automation}
\label{section:ui_interaction_automation}
Many researchers also investigated the potential of AI to automatically interact with UIs, which reduced the user efforts required for creating task automation compared with e.g., programming by demonstration and interactive task learning methods~\cite{li_pumice_2019, li_sugilite_2017,li_interactive_2020}. Over the years, researchers have explored single-turn, UI-element-based simple interactions~\cite{degott_learning_2019, todi_conversations_2021, wang_enabling_2023}, to multi-turn~\cite{iki_berts_2022, yao_webshop_2023, furuta_instruction-finetuned_2023, wen_empowering_2023}, more complex and precise actions, such as horizontal scrolls~\cite{rawles_android_2023}. Language models~\cite{todi_conversations_2021, iki_berts_2022, wang_enabling_2023, furuta_instruction-finetuned_2023, rawles_android_2023, wen_empowering_2023} and reinforcement learning~\cite{degott_learning_2019, yao_webshop_2023} are the most utilized approaches for predicting action sequences in UIs. The understanding and prediction of user actions on UI screens can support diverse downstream designer-centric tasks, such as facilitating the prototyping of user flows, simplifying existing user experiences, and understanding user goals and intents.

\subsubsection{Summary}
In all, papers in this section focus on exploring and extending AI models' general abilities on tasks related to UI and UX. The advancement of these AI models' capabilities can benefit from ongoing innovations in the AI community. Multi-modal model pipelines remain mainstream with UI/UX datasets and have continuously demonstrated impressive performance~\cite{zhang_screen_2021, li_screen2vec_2021, wang_screen2words_2021, ang_learning_2022}. Compared to traditional computer vision tasks, merely considering pixel information is far from satisfactory. Other modalities, such as structural information in vector graphics, are quite common in human design practices and thus require more attention in these model architectures.
Recently, the significant increase in AI model sizes, reflected in pipelines based on BERT~\cite{bai_uibert_2021} and large vision-language models~\cite{li_spotlight_2023} provide potential future directions. Meanwhile, deeper engagement with the UX communities and human-centered approaches can help uncover more direct translational opportunities to support UX practitioners with AI, as discussed in Section~\ref{section:dataset-discussion}. We believe these two complementary approaches are both indispensable in pushing forward the boundary of AI-driven UX design support tools.

%% file: sections/5-discussion.tex
\section{Discussion}
\label{section:discussion}

In the previous section, we have mapped the existing literature on AI's role in UX support, applying the Double Diamond framework to structure our exploration. Here, from a meta-level perspective, we draw inspiration from existing Human-Centered AI research~\cite{horvitz_principles_1999, amershi_guidelines_2019, lubars_ask_2019, shneiderman_human-centered_2022} and distill the patterns observed across all phases, aiming to summarize and discuss generalizable insights. These insights pinpoint details of the gap between technical AI research and the human-centered UX mindset, emphasizing the need for the collaborative adaptation and evolution of both domains to better complement each other.

\subsection{AI Assistance for UX: A Promising Field for Interdisciplinary, Translational Research}

Our systematic literature review has demonstrated that the area of AI assistance for UX has witnessed significant growth. Research across HCI and AI has pushed the boundaries of AI datasets and models for UI/UX, understanding UX practices, and applications of technical innovations into various design activities.

Various AI techniques have been effectively utilized to process, understand, and generate user interface data, an inherently rich, multimodal data format~\cite{deka_rico_2017, rawles_android_2023}. Techniques and methodologies from subfields of AI, such as natural language processing, computer vision, graph learning, and reinforcement learning, have all been utilized, often in combination, to experiment with UI datasets~\cite{liu_reinforcement_2018, zhang_screen_2021, wang_screen2words_2021, li_screen2vec_2021, schoop_predicting_2022, wang_learning_2020, eskonen_automating_2020, bruckner_learning_2022, hotti_graph_2022}. Research in the field has been consistently reflecting AI breakthroughs, with the latest adoption being Large Language Models~\cite{wang_enabling_2023} and Large Vision-Language Models~\cite{li_spotlight_2023}. To this extent, UX research and design have provided AI researchers with unique challenges to tackle, effectively benefiting the AI community.

UX research and design are also fertile grounds for translational research~\cite{colusso2017translational, norman2010research, colusso2020understanding} to impact the UX industry. It also has generalizable values for various adjacent domains. For example, the non-linear aspect of UX processes makes the research findings generalizable to AI-supported creativity research. Future research will continuously provide immense opportunities for both AI and UX to collaboratively evolve.

\subsection{Unique Characteristics of UX: Empathy Building and Emphasis on Experiences}

\subsubsection{The Essence of UX Methodologies: Empathy Building}

A central goal of UX methodologies and processes is \textit{empathy building}. Previous research has uncovered that UX practitioners view methodologies more as ``\textit{mindsets}'', rather than actual rigorous methods, to scaffold listening to users and considering diverse user inputs~\cite{gray_its_2016}. Practitioners emphasized the need to prioritize this mindset when \textit{adopting} and \textit{adapting} UX methodologies based on each project's unique scenarios:

``\textit{... methods themselves are quite rudimentary... you probably can describe in a page.  But when it comes to actually getting the right value out of them, it’s having that right mindset –– what are the right questions we need to ask? How can we
answer them? And then using that as the basis for what
methods you need.}''~\cite{gray_its_2016}

This unique characteristic of UX is often overlooked in current AI support tools: it is often \textbf{\textit{not}} about automating the processes or methodologies, but supporting UX designers' and researchers' empathy-building with their users. Simplistic automation and acceleration can overlook the cognitive process of UX professionals, thus obstructing the researchers’ learning and empathy-building~\cite{marathe_semi-automated_2018}. As a result, existing research that uses AI for simplistic automation, as we have discussed in Sections~\ref{section:discover} and ~\ref{section:define-phase}, is generally not desired by UX practitioners and hard to integrate into existing workflows. For example, AI systems that aim to directly provide synthetic user information~\cite{zhang_data-driven_2016, an_imaginary_2018, tan_generating_2022} can obstruct the empathy-building goal of UX and reinforce stereotypes by providing ``statistically most likely'' information about users~\cite{salehi_i_2023}. Addressing this gap calls for more adoption of a human-centered AI perspective~\cite{shneiderman_human-centered_2022}, orienting future research to support the \textit{mindsets} and \textit{goals} of UX designers, instead of simply automating UX processes and generating relevant UI/UX artifacts.

\subsubsection{From Individual UI Screens To Underlying User Experiences}
\label{section:beyond-individual-screens}
Most of past research focused on individual UI screens, often overlooking the user flows and user experiences encapsulated across multiple interfaces. As we discovered in Section~\ref{section:datasets}, most existing datasets focused on static UI screens and components. AI models and their applications, built on top of these datasets, also mostly worked with static UIs (see Sections~\ref{section:develop}, ~\ref{section:deliver},~\ref{section:general-ai-models}). Only a limited number of exceptions were found in our sample, focusing on topics such as user engagement~\cite{wu_predicting_2020} and creating UI animations~\cite{natarajan_p2a_2018}. 

This is a main factor that limits existing research's practical application in the real-world UX industry. During the past years, there has been a notable shift of design's focus, from user interfaces to holistic user experiences~\cite{norman1998definition}. This shift has been further amplified by the wide adoption of design systems~\cite{churchill2019scaling}, i.e. libraries of UI components and styles defined within companies to ensure consistent visual styles and branding across products (e.g. Google Material Design, IBM Carbon, Microsoft Fluent). The high-fidelity design components in design systems are defined with great detail and precision. Thus, designers are constrained in changing the visual aspects of design, while freed to focus more on crafting friendly, seamless user experiences with pre-defined UI elements~\cite{frost_atomic_2016}. Consequently, this gap between academic explorations and industry practices widens, limiting practical, real-world adoption of AI-enabled design support tools created in academic settings.

However, great potentials for AI support still exist, if coupled with a deep understanding of existing UX practices and workflows. Recent research in UI task automation, from datasets like Android in the Wild~\cite{rawles_android_2023} to models such as UIBert~\cite{bai_uibert_2021}, reflect a gradual shift from individual UI screens to the underlying user flows and experiences. While the users' perspective of interacting with UIs can still be different from designers' considerations, great opportunities for design support tools lie in designer-centric applications of these datasets and AI models. In addition, as UI animation and motion design increasingly become integral parts of modern user experience~\cite{motion_material}, video-based AI models present promising avenues to enhance relevant design processes and tools~\cite{wu_predicting_2020, natarajan_p2a_2018}.

\subsection{Analyze Task Delegability in UX Workflows}

Given the intricacies of UX processes and methodologies, it is necessary to consider UX practitioners' goals when determining the delegability of tasks to AI. Delegability is a concept in human-centered AI that describes the extent to which AI should be involved in certain tasks~\cite{jiang_supporting_2021, feuston_putting_2021}. Lubars and Tan proposed a framework of task delegability for AI, considering the motivation, difficulty, risk, and trust when deciding the involvement of AI~\cite{lubars_ask_2019}. UX processes are often fluid and non-linear yet tied to practical business and design goals (e.g. higher conversion rate, increased user engagement)~\cite{li2024understanding}. Such processes blend both creative and analytical tasks, complicating AI task delegability analysis.

In the context of UX, our analysis has highlighted UX practitioners' main motivation for empathy-building with users. We encourage future researchers to carefully analyze the different UX methodologies, as well as the detailed steps within these methodologies, against the task delegability framework. The level of AI automation and the granularity of task breakdown are not binary choices, but balances to be kept when designing UX support tools~\cite{shneiderman_human-centered_2022}. For example, for qualitative analysis (Section~\ref{section:define-phase}), prior research suggests less AI delegability for the initial codebook creation than for the later codebook labeling process~\cite{marathe_semi-automated_2018}. Systems like Cody~\cite{rietz_cody_2021} and PaTAT~\cite{gebreegziabher_patat_2023} serve as positive examples, where AI models automate manual work but still leave room for reflections, learnings about users, and empathy-building. The appropriate amount of automation with AI in suitable tasks will improve the quality of UX outcomes, helping practitioners get the right design and get the design right.

\subsection{Designer-Centric Datasets and Evaluation Metrics as Solid Technical Foundations}

Most of the current open source UI/UX datasets~\cite{deka_rico_2017, moran_machine_2020, leiva_enrico_2020} and user task automation~\cite{rawles_android_2023} are not directly associated with designers' considerations and priorities when designing for user experiences. Even the datasets in our sample that are created for design purposes are often limited to individual UI screens and widgets~\cite{huang_swire_2019, sermuga_pandian_uisketch_2021, sermuga_pandian_synz_2021}, ignoring practitioners' emphasis on user experience across screens. For example, UX designers' key considerations in the design process can include: how to identify appropriate design patterns for a given design scenario~\cite{silva-rodriguez_machine_2019}, how to implement user flows and product features with existing UI components from a design system~\cite{churchill2019scaling}, and how to fit UI components into existing screens to support additional product features~\cite{lu_bridging_2022}.

This calls for future research contributions in two main areas: first, we need better evaluation metrics of UI/UX data that align more closely with current UX design goals, such as usability heuristics~\cite{ponce_deep_2018} (also see Section~\ref{section:discussion-2}); second, more datasets containing the results of such metrics are needed for large-scale benchmarking efforts. Metrics that effectively reflect the quality of UI/UX design are still not efficient, which nowadays is mostly achieved through subjective scores provided by potential user groups~\cite{swearngin_rewire_2018, ang_learning_2022}. UX designers and researchers should work closely with AI researchers and engineers in defining these metrics, as well as contributing to such data collection and labeling efforts~\cite{yildirim_how_2022}.

The lowering barrier of utilizing AI, through pre-trained large language models~\cite{wang_enabling_2023} and large vision-language models~\cite{li_spotlight_2023}, can also enable more UX teams to easily integrate AI into their toolkits~\cite{brie_evaluating_2023, xiao_supporting_2023, feng_layoutgpt_2023}. Interacting with these models, whether through fine-tuning or prompt-based mechanisms, reduces the reliance on domain-specific datasets. This is advantageous for UX professionals leaning towards qualitative methods or those lacking the means to collect large-scale datasets. However, employing such extensive models also raises ethical concerns over potentially leaking user data, emphasizing the need for more ethical research on the responsible deployment of these models~\cite{shen_shaping_2023}.

%% file: sections/6-conclusion.tex
\section{Conclusion}
This study underscores the expanding potential of integrating AI into the UX domain through a systematic literature review (SLR) from a Human-Centered AI (HCAI) perspective.By mapping research onto the Double Diamond framework, we identified key technical capabilities of AI in UX and highlighted overlooked aspects like empathy-building and multi-screen user experiences. We highlight the need for a deep understanding of UX practices, mindsets, and goals to design effective AI support, calling for careful AI delegability analysis for UX tasks based on existing HCAI frameworks~\cite{lubars_ask_2019}. Designer-centric datasets and evaluation metrics can greatly improve the technical foundations for direct real-world impact. This review summarizes the current landscape and lays out future opportunities for this promising interdisciplinary, translational research domain.

%% file: sections/7-limitations.tex
\section{Limitations}

Given the rapid advancements in our focus areas, keeping up-to-date with the latest developments is challenging, particularly for a systematic literature review. Our snowball sampling method enabled us to gather a substantial set of relevant papers (N=369), but this approach limits our ability to incorporate new studies as they are published. Notably, recent papers on UI task automation using LLMs~\cite{li2023zero, yan2023gpt} appeared after our review period and were not included in our analysis. Despite this, we believe our findings remain relevant and insightful in light of these new publications. Nonetheless, we acknowledge this as a limitation of our study.

%% file: sections/z-appendix.tex
\section{Appendix}
\label{sec:appendix}

\subsection{Rationale to Use Snowball Sampling}

\begin{table*}[ht]
\centering
\begin{tabular}{p{8cm}|c|c}
\hline
\textbf{Query} & \textbf{ACM DL} & \textbf{Google Scholar} \\
\hline
("UX" or "user experience") and ("AI" or "Artificial Intelligence" or "ML" or "Machine Learning") & 292,808 & 2,580 \\
\hline
("UI" or "user interface" or "UX" or "user experience" or "HCI") and ("AI" or "Artificial Intelligence" or "ML" or "Machine Learning") & 67,164 & 5,510,000 \\
\hline
("UI design" or "UX design" or "interaction design") and ("AI" or "ML") & 3,222 & 16,900 \\
\hline
\end{tabular}
\caption{Search Queries and Results}
\label{tab:search_results}
\end{table*}

As discussed in section~\ref{sec:method}, we used snowball sampling for literature selection instead of keyword-search in popular search engines or databases. We initially attempted to use keyword search and tested out a few advanced queries in ACM Digital Library and Google Scholar. A few advanced search queries about "UX" and "AI", or "machine learning" in these systems all resulted in hundreds of thousands of papers, similar to~\cite{stefanidi_literature_2023}. Most of the search results were not relevant to our scope and it is hardly practical for us to manually filter through them all. Moreover, these online search engines/databases often disagree with each other in their results significantly, posing uncertainty regarding their reliability. As a result, we conducted snowball sampling instead. Here we attach a few queries we attempted to use and the count of their search results as of August 2023 when we write this paper.